\documentstyle[a4,11pt,epsf]{article}
\epsfverbosetrue
 0

\newcommand{\beq}{\begin{equation}}
\newcommand{\eeq}{\end{equation}}

\begin{document}
             

\begin{titlepage}

\vskip 1.5 cm
\begin{center}
{\huge \bf 
Critical phenomena:\\[0.5cm]
 150 years since Cagniard de la Tour}
\end{center}

\vskip 2.0 cm
\centerline{ {\bf Bertrand Berche}$^a$, {\bf Malte Henkel}$^a$ and {\bf Ralph Kenna}$^b$}
\vskip 0.5 cm
\centerline {$^a$D\'epartement de Physique de la Mati\`ere et des Mat\'eriaux, Institut Jean Lamour,\footnote{Laboratoire associ\'e au CNRS UMR 7198} 
} 
\centerline{CNRS -- Nancy Universit\'e -- UPVM, B.P. 70239,}
\centerline{F -- 54506 Vand{\oe}uvre l\`es Nancy Cedex, France}

\centerline{$^b$Applied Mathematics Research Centre, Coventry University,}
\centerline{Coventry CV1 5FB, England} 

\begin{abstract}
Critical phenomena were discovered by Cagniard de la Tour in 1822, who died 150 years ago. In order to mark this anniversary, the context and the early history of his discovery is reviewed. We then follow with a brief sketch of the history of critical phenomena, indicating
the main lines of development until the present date. \\~\\

Os fen\'omenos cr\'{\i}ticos foram descobertos pelo Cagniard de la Tour em Paris em 1822. Para comemorar os 150 anos da sua morte, o contexto e a hist\'oria initial da sua descoberta \'e contada. Conseguimos com uma descri\c{c}\~ao breve da hist\'oria dos fen\'emenos cr\'{\i}ticos, indicando as linhas principais do desenvolvimento at\'e o presente. 
\end{abstract}

\end{titlepage}

BARON CHARLES CAGNIARD DE LA TOUR (1777 -- 1859), who died 150 years ago, 
was the discoverer of critical phenomena. 
What started as an  exotic curiosity has developed into a mature field
underlying the modern-day physics of many-body and complex systems.
Here the circumstances surrounding his discovery are reported, and the 
evolution of the field to the present-day is synopsised.

\begin{figure}[t]
\centerline{\epsfxsize=4.0cm\epsfbox{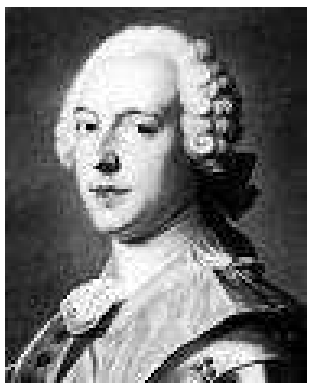}}
\caption{Portrait of Cagniard de la Tour, courtesy of the Universidade do Minho, Portugal. \label{fig:Cagniard1} }
\end{figure}

\begin{figure}[h]
\centerline{\epsfxsize=7.0cm\epsfbox{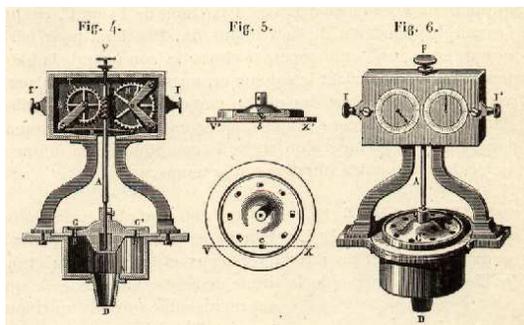}}
\caption{The improved siren, invented and named by Charles Cagniard de la Tour. Photograph courtesy of the {\'{E}}cole Polytechnique Paris, France. \label{fig:Cagniard2} }
\end{figure}

Born in Paris in 1777,  Charles Cagniard
was educated at l'{\'{E}}cole Polytechnique, and went on to become a 
prolific scientist and inventor. 
Besides his discovery of critical phenomena, 
Cagniard de la Tour investigated the nature of yeast
and its role in the fermentation of alcohol and was interested the physics of the 
human voice as well as bird flight.
His interest in  acoustics led to the invention of the siren (see figure~\ref{fig:Cagniard2}), 
which he named after sea creatures from Greek mythology who lured sailors to their
doom.

Experiments on steam engines in the late 17th and early 18th centuries motivated 
interest in the behaviour
of fluids at high temperatures and pressures. Denis Papin (1647 -- 1712)
who invented the \emph{``steam digester''} -- a forerunner of the steam 
engine -- noticed that when heated under 
pressure, water remains in its liquid phase at temperatures far 
greater than the usual boiling point of 100$^\circ$C: the temperature of the
boiling point increases with increasing pressure. 

The term \emph{``latent heat''}, for the energy required to complete a 
solid-liquid or liquid-vapour phase transition, was introduced around 1750 by 
Joseph Black (1728 -- 1799).
In 1783  James Watt (1736 -- 1819) analysed its dependency on pressure, and
found that the latent heat of vaporisation decreases as the temperature is increased. 
At this time, gases were considered to be distinct from vapours (produced by evaporating liquids). 
\emph{``Elastic fluids''} which were not reducible  to liquid form were termed gases. 
It was in the second half of the 18th century
that  Antoine-Laurent de Lavoisier (1743 -- 1794) 
showed gases and vapours to be one and the same, and a third state 
of matter beside solids and liquids.
He also suggested that gases could be liquefied at sufficiently low 
temperature and high pressure \cite{Lavoisier}.

The first successful experiments on liquefaction of gases took place in 1784, 
when  Jean-Fran\c{c}ois
Clouet (1751 -- 1801) and Gaspard Monge (1746 -- 1818) achieved the 
liquefaction of gaseous sulphur dioxide  by cooling and compression. 
There followed a sequence of successful experiments, including by  chemist 
and physicist Michael Faraday (1791 -- 1867),  in which gases were liquefied,
thus removing the distinction between vapour and gas \cite{Fa1823,Fa1824}. 
Hydrogen, oxygen, nitrogen, and carbon monoxide, which  were previously thought 
to be incondensably  gaseous and were called \emph{``permanent gases''} were 
eventually liquefied 1877.

The discovery of what we now call the critical point came about with 
Cagniard de la Tour's experiments with Papin's digester.
In 1822,  in the context of his interests in acoustics, he placed a flint
ball in a  digester partially filled with liquid.
Upon rolling the device, a splashing sound was generated as the solid ball penetrated the 
liquid-vapour interface.
Cagniard de la Tour noticed that upon heating the system far beyond the 
boiling point of the liquid,  
the splashing sound ceased above a certain temperature.
This marks the discovery of the supercritical fluid phase. 
In this phase there is no surface tension as there is no liquid-gas phase boundary. 
The supercritical fluid can dissolve matter like a liquid and can diffuse 
through solids like a gas. 

\begin{figure}
\centerline{\epsfxsize=12.0cm\epsfbox{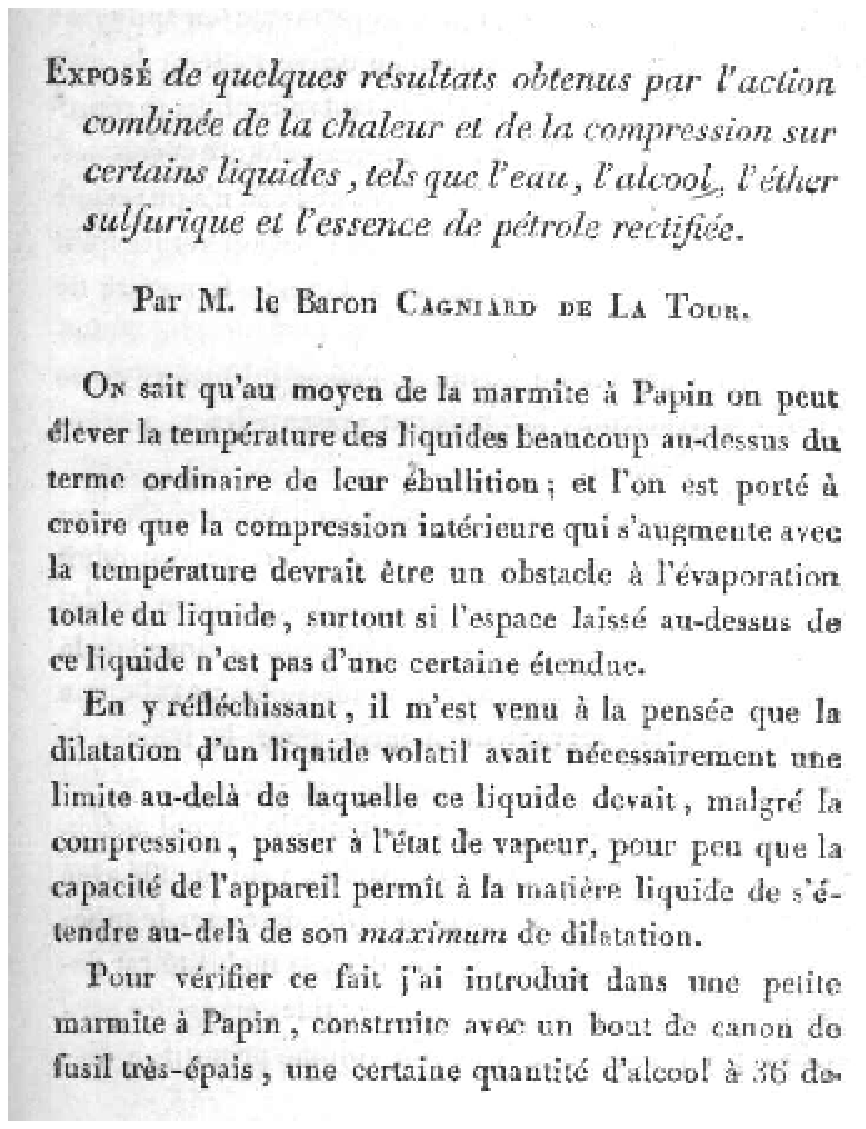}}
\caption{The first page of Cagniard de la Tour's article, in which the discovery of critical phenomena is reported.\label{fig:Cagniard3} }
\end{figure}

In two articles in the {\emph{Annales de Chimie et de Physique}} \cite{LaTour1822},
Cagniard de la Tour described how he heated a sealed glass tube of alcohol under pressure,
see figure~\ref{fig:Cagniard3}. 
He  observed that the liquid expanded to approximately twice its original volume,
and then vanished, having been converted to a vapour so transparent that the tube 
appeared completely empty.
On re-cooling the system a thick cloud appeared. 
We now recognise this as an observation of critical opacity and the discovery 
of the critical point.
He also observed that beyond a certain temperature, increasing the pressure did not prevent the evaporation of the liquid.

In a following paper, Cagniard de la Tour reported upon a series of 
related experiments with a variety of substances \cite{LaTour1823}.
Desiring to demonstrate that the existence of a limiting temperature
above which a liquid vapourises irrespective of pressure
is a general phenomenon,
he experimented on water, alcohol, ether and carbon bisulphide.
He measured the critical temperature at which the interface tension vanished, 
as determined by the 
disappearance of the meniscus, and
discovered that for each substance, there is a certain temperature beyond which  
total vaporisation of the liquid occurs and where 
no increase in pressure will liquefy the gas. In the case of water,
 this critical temperature was estimated to be 362$^\circ$C,
a remarkably accurate result (modern measurements give 374$^\circ$C).
His experiments demonstrated that this {\emph{``{\'{e}}tat particulier''}} 
requires high temperatures, almost independent of the volume of the 
tube: ``{\emph{... cet {\'{e}}tat particulier exige toujours une temp{\'{e}}rature
tr{\`{e}}s-{\'{e}}lev{\'{e}}e, presque ind{\'{e}}pendante de la capacit{\'{e}} du tube}\/}''
\cite{LaTour1823}.
We now know that the  {\emph{{\'{e}}tat particulier}} marks the critical 
end-point of a line of first-order phase transitions, where the transition becomes continuous.

While many of Cagniard de la Tour's contemporaries regarded his results as being particular 
to the substances involved rather than a general 
phenomenon \cite{Go94},  Faraday recognised the significance of his work \cite{Fa1824}.
In a letter to William Whewell in 1844, Faraday wrote \cite{FaradayLetter1844}
{\it ``Cagniard de la Tour made an experiment some years ago which gave 
me occasion to want a new word''}.  
Referring to what we now call the critical point, he continued, 
{\it ``how am I to name this point at which the fluid \& its vapour 
become one according to a law of continuity.
Cagniard de la Tour has not named it; what shall I call it?''}
Whewell suggested to call it the point of {\emph{vaporiscience}} 
or the point at which fluid is {\emph{disliquified}}
or the {\emph{Tourian state}}, 
and in a later publication  Faraday refers to  
{\emph{``Cagniard de la Tour's state''}} and {\emph{``the Cagniard de la Tour point''}}  \cite{Fa45}.
In 1861, Dmitri Mendeleev  (1834 -- 1907), 
referred to it as the  
{\emph{``absolute Siedetemperatur''}\/}, or {\emph{absolute boiling point}}
 \cite{Me1861}.

In 1869, the term  we now use -- the {\emph{critical point}} --  was eventually coined
by Thomas Andrews (1813 -- 1885), who 
further elucidated the meaning of Cagniard de la Tour's 
{\emph{{\'{e}}tat particulier}\/} \cite{Andrews}. 
Andrews studied the pressure-volume curve of the liquid-vapour  coexistence 
line of carbonic acid 
and clarified that a gas may only condense to a liquid, or a liquid evaporate 
to a gas, below certain values of the
temperature and pressure -- the {\emph{{\'{e}}tat particulier}}. Beyond this point lies the 
supercritical phase, where the distinction between liquid and vapour disappears.

In what followed, the early experiments of Cagniard de la Tour blossomed
into a large-scale intellectual adventure. 
In 1873, van der Waals (1837 -- 1923) showed in his doctoral thesis \cite{vdW} 
that Andrews' experimentally based equation of state may be 
explained qualitatively using  an extension of the ideal gas law which modelled molecular 
attraction and hard-core repulsion in a simple manner. This in turn suggested to
Heike Kamerlingh Onnes (1853 -- 1926) how to estimate the critical points for `permanent gases', 
which gave the conceptual bases for the eventual liquefaction of helium, followed soon after by the
discovery of superconductivity. On the other hand, the simple mean-field-like values 
of the \emph{``critical exponents''} obtained from his equations are not adequate for a
quantitative description of real systems, as realised experimentally in 1896 by
Jules-{\'{E}}mile Verschaffelt (1870 -- 1955). Mean-field-like treatments were 
systematised in the phenomenological theory of Lev Davidovich Landau (1980-1968), 
where phase transitions in all spatial dimensions were predicted \cite{La37}.

On the other hand, the important concept of \emph{``universality''} of critical 
phenomena was introduced by Pierre Curie (1859 -- 1906), 
who discovered that ferromagnetic materials become demagnetised above a critical temperature
 \cite{Curie1891}
which is often referred to as a \emph{``Curie point''}. 
Formal analogies between {\it a priori} unrelated physical systems have been of great
usefulness in trying to understand critical phenomena and were also 
one of the motivations when Wilhelm Lenz (1888 -- 1957) 
introduced the simple many-body system now usually called
\emph{``Ising model''} \cite{Le20}. 
Ernst Ising (1900 -- 1998) solved  the one-dimensional case 
in his doctoral thesis (1924) and the absence
of a phase transition there clearly showed that a conceptual explanation of the 
critical point beyond the level of
mean-field theories had to be sought. This conclusion was further strengthened
by the achievements of Lars Onsager (1903 --1976), who in 1944 calculated  exactly the
specific heat of the two-dimensional Ising model in the absence of an external
magnetic field and in 1949 announced the correct
formula for the spontaneous magnetisation, proven by C.N. Yang (1922 -- ) in 1952. 
By a tour de force and combining techniques of conformal field-theory with 
integrable systems, 
Alexander Zamolodchikov (1952 -- ) showed in 1989 that the two-dimensional 
Ising model in an external
magnetic field, but with the temperature fixed to the critical temperature, 
is integrable \cite{Zamo89}. 

In view of the absence of an exact solution for the three-dimensional Ising model, 
numerical techniques came to the fore. These are based either on systematic 
expansions around the known extreme 
cases of very high or very low temperatures as suggested by Cyril Domb (1920 -- ) 
in his doctoral thesis in 1949 \cite{Domb96}, or else are based on large-scale simulations 
which go under the name of \emph{``Monte Carlo method''} and suggested in 1949 
by Nicholas Metropolis (1915 --1999) and Stanislaw Ulam (1909 -- 1986) \cite{MetroMC}. 
In the 1960s it was realised  
by Leo Kadanoff (1937 -- ) and Michael Fisher (1931~--~) that a general theoretical 
framework for phase transitions would have to be formulated in terms
of a \emph{``scaling theory''} which in particular led to \emph{``scaling relations''}
between the critical exponents which describe the behaviour of the various measurable
quantities close to a critical point. 
This opened the way to a full theoretical description of
critical phenomena through the 
\emph{``renormalisation group''} by Kenneth Wilson (1936 -- ) in 1971. This has been
the basis for very precise predictions of the values of the critical exponents
in two and three dimensions. 

On the other hand, since the days of Cagniard de la Tour experimental techniques
have been continuously refined. Very precise estimates for the values of the
critical exponents can nowadays be obtained. 
For a long time, however, while experimentalists were busy measuring the
critical behaviour of three-dimensional bulk systems, theorists could only calculate exactly the critical behaviour of two-dimensional systems, which 
can only be realised at the {\it surface} of some substrate. 
It took surprisingly long until phase transitions for systems confined to a 
surface were experimentally observed. The first confirmed example seems to 
have been found in Nancy by Andr\'e Thomy in his th{\`{e}}se 3{\`{e}}me 
cycle (1959) for krypton adsorbed on graphite \cite{Thomy70}. 
Fittingly, this discovery arose exactly a century after the death of
Cagniard de la Tour.

Nowadays, the most precise experiments are carried out on board
the space shuttle, the space station MIR and the International Space Station. 
As an example, we quote the result for the specific-heat exponent 
$\alpha = 0.11\pm 0.03$  obtained for the critical point in the simple 
fluid SF$_6$ during the Spacelab D2 mission (1999), 
in good agreement with the current theoretical 
estimate $\alpha=0.109 \pm 0.002$ \cite{Barmatz07}. 
(Mean-field theory would have predicted $\alpha=0$.) 

In the 150 years since its inception, the field of critical phenomena 
has blossomed and now forms a cornerstone of modern physics, 
both experimental and theoretical and this development nicely illustrates how a topic of purely fundamental research, given enough time, can diversify into initially unforeseeable directions. 
Its founder, Charles Cagniard de la Tour died in Paris on the 5$^{\rm th}$ of July 1859.

\bigskip
%

\end{document}